\documentclass[final,5p,twocolumn]{elsarticle}

\usepackage{booktabs}
\usepackage{amssymb}
\usepackage{amsmath, amsfonts}
\usepackage{underscore}
\usepackage{multirow}
\usepackage{multicol}
\usepackage{float}
\usepackage{stfloats}
\usepackage{flushend}
\usepackage{wrapfig}
\usepackage{graphicx}
\usepackage{mwe}
\usepackage{subcaption}
\usepackage[table]{xcolor}
\usepackage[nameinlink,capitalise]{cleveref}
\usepackage{lineno}\modulolinenumbers[5]
\usepackage{blindtext}
\usepackage{changepage}
\usepackage[export]{adjustbox}
\usepackage{array}
\usepackage[utf8]{inputenc}

\renewcommand\toprule{\specialrule{1pt}{1pt}{1pt}\rowcolor{gray!10}}
\renewcommand\midrule{\specialrule{0.4pt}{0.4pt}{3pt}}

\newcommand\doubleRulee{\specialrule{1.2pt}{1pt}{0.5pt}\specialrule{0.5pt}{0.5pt}{2pt}}

\journal{Engineering Science and Technology, an International Journal}

\makeatletter
\def\ps@pprintTitle{%
	\let\@oddhead\@empty
	\let\@evenhead\@empty
	\def\@oddfoot{\footnotesize\itshape
		Preprint submitted to Engineering Science and Technology, an International Journal\hfill}%
	\let\@evenfoot\@oddfoot}
\makeatother

\begin{document}
	
	\begin{frontmatter}
		
		\title{An Innovative Data Collection Method to Eliminate the Preprocessing Phase in Web Usage Mining\tnoteref{tref}}
		\tnotetext[tref]{This work was supported by Sakarya University Scientific Research Foundation (Project number: 2010-50-02-024).}
		
		\author[a,b]{Özkan Canay\corref{cor1}}
		
		\affiliation[a]{organization={Sakarya University},
			addressline={Institute of Natural Sciences, Dept. of Computer and IT Engineering},
			city={Serdivan},
			postcode={54050},
			state={Sakarya},
			country={Turkiye}}
		\affiliation[b]{organization={Sakarya University of Applied Sciences},
			addressline={Vocational School of Sakarya, Dept. of Computer Tech.},
			city={Adapazari},
			postcode={54290},
			state={Sakarya},
			country={Turkiye}}
		\cortext[cor1]{Corresponding author}
		\ead{canay@subu.edu.tr}
		\ead[url]{https://canay.subu.edu.tr/}
		
		\author[c]{Ümit Kocabıçak}
		\affiliation[c]{organization={Sakarya University},
			addressline={Faculty of Computer and IT Engineering, Dept. of Computer Eng.},
			city={Serdivan},
			postcode={54050},
			state={Sakarya},
			country={Turkiye}}
		
		\begin{abstract}
			The underlying data source for web usage mining (WUM) is commonly thought to be server logs. However, access log files ensure quite limited data about the clients. Identifying sessions from this messy data takes considerable effort, and operations performed for this purpose do not always yield excellent results. Also, this data cannot be used for web analytics efficiently. This study proposes a method for user tracking, session management, and collecting web usage data. The method is mainly based on an innovative approach for using collected data for web analytics extraction as the data source in web usage mining. An application-based API has been developed with a different strategy from conventional client-side methods to obtain and process log data. The log data has been successfully gathered by integrating the technique into an enterprise web application. The results reveal that the homogeneous data collected and stored with this method is more convenient to browse, filter, and process than web server logs. This structured data can be used effortlessly as a reliable data source for high-performance web usage mining activity, real-time web analytics, machine learning algorithms, or a functional recommendation system.
		\end{abstract}
		
		
		\raggedbottom
		\begin{highlights}
			\item The proposed method allows the collection of visitor web usage data.
			\item The method is an alternative to server logs used as a data source for web usage mining.
			\item Guarantees organizations complete control and ownership of their data.
			\item Works faster than the client-side methods and makes the server less busy.
			\item The obtained data is suitable for multi-purpose use.
		\end{highlights}
		
		\begin{keyword}
			Web usage mining \sep Preprocessing \sep Data collection \sep Log analysis \sep Web analytics
		\end{keyword}
		
	\end{frontmatter}
	
	
	\section{Introduction}\label{sec:1}
	The rapid development and widespread use of information technologies have contributed enormously to the evolution of the web. As a natural consequence of this mutually nourishing cycle, organizations have shifted their business processes and efforts to this media. Information is the most crucial asset for many organizations, and the web is an immense data source. Those who are aware of this shift are investing more than ever in collecting and making sense of more data to enhance their business strategies. To know their visitors and behavior well, particularly for organizations that position their focus on the web, is of paramount importance.
	
	Most web-based applications such as corporate web applications, web portals, shopping sites, or learning management systems keep logs that focus on user transactions, albeit relatively superficial. However, for organizations that approach the issue from a strategic point of view, understanding visitor behavior, recognizing the user profile, and shaping their actions according to this requires much more effort than ordinary logging.
	
	Web servers accumulate all page requests in access logs unless otherwise stated. Every click from a search engine, every page visit, or download is recorded continuously, anonymously, and silently. These access logs, which contain identities, origins, and browsing behaviors, are preferred in scientific studies because they provide ready-made data \cite{ref1}.
	
	The log files of user-intensive websites grow by hundreds of megabytes daily. This situation makes analysis time-consuming and places essential performance requirements on the HTTP log analyzer tools. The most common complaints about the analyzers are that they are not flexible enough, have limited availability, are challenging to use, and are extremely slow, especially on big data \cite{ref2}.
	
	On the other hand, while third-party web analytics tools present calculated analytics information, they do not provide raw data on which to perform WUM processing. Although there are many publications on web analytics and web usage mining in the literature, there is no study suggesting the use of these two processes together. Moreover, since web analytics is often a commercial activity, companies do not prefer to publish scientifically how these tools collect and store data.
	
	Cloud-based analytics tools usually use client-side user tracking methods and store data on remote servers. These tools do not provide downloading raw data since it is enormous in size and stored in the data warehouse. Therefore, it is not possible to apply web mining techniques to this data. Over and above, although such tools are free for everyday use, quotas ensured are insufficient for organizations with a large user base. In contrast, paid versions are quite expensive, and still, they do not allow downloading raw data.
	
	Cloud-based web analytics tools are also an essential source of data privacy and sovereignty concerns. The open-source tools such as Piwik, OWA \cite{ref3}, and Webalyt \cite{ref4}, which have emerged to address these concerns, store the data on local servers. These tools use client-side JavaScript snippets for user tracking, just like commercial alternatives. The working principles of almost none of these tools have been explained sufficiently in the literature. In addition, these tools, which conduct the user tracking process on the client side, increase the load on the server by performing a separate request for each page transition.
	
	\begin{table*}[b]
		\centering
		\caption{Recent studies about WUM preprocessing and web analytics fields}
		\label{tab:1}
		\small
		\begin{tabular}{ m{0.60\textwidth} >{\centering\arraybackslash}m{0.035\textwidth} m{0.17\textwidth} >{\centering\arraybackslash}m{0.025\textwidth}  }
			\toprule
			Key Contribution & Year & Technique	& Ref.\\
			\midrule
			
			Extracting data from log files in different subsystems in respective domains & 2020 & Log data fusion & \cite{ref43}\\
			A presented model for path analysis based on full paths & 2019 & Path analysis & \cite{ref44}\\
			Detailed preprocessing to improve the quality and effectiveness of weblog data & 2019 & Preprocessing & \cite{ref45}\\
			Two algorithms presented for analyzing visitor behavior and preprocessing & 2020 & Preprocessing & \cite{ref46}\\
			A MapReduce-based data algorithm for all stages of preprocessing & 2021 & Preprocessing & \cite{ref47}\\
			A graph-based method for session construction in preprocessing & 2022 & Session identification & \cite{ref48}\\
			Using entropy in preprocessing for session identification & 2018 & Session identification & \cite{ref49}\\
			MapReduce-based user identification algorithm in web usage mining & 2018 & User identification & \cite{ref50}\\
			Using AWStats to analyze public logs  & 2017 & Web analytics & \cite{ref51}\\
			A new open web analytics platform proposal, Webalyt & 2017 & Web analytics & \cite{ref4}\\
			Using Matomo for open government & 2021 & Web analytics & \cite{ref52}\\
			\hline
		\end{tabular}
	\end{table*}

	Web usage mining and web analytics are sharply separated, although both use web access data. Web analytics involves collecting log data and presenting them visually by statistical extraction. In contrast, web usage mining focuses on identifying valid, novel, helpful, and understandable patterns using knowledge discovery techniques after cleaning and structuring existing data. Therefore, the WUM assumes that this data is already available; it is not interested in acquiring or accumulating them. 
	
	A novel method is proposed in this study as an alternative to web server logs used as WUM data sources. The method eliminates preprocessing phase, the most demanding step of the WUM process, and provides much more reliable and convenient data from server access records. It also allows the collection and processing of visitor usage data at the application level.
	
	The innovative aspect of this study is that it proposes a data collection method for organizations aiming to develop their own web analysis systems and demonstrates the usability of the data collected by this method for web usage mining. The data obtained by the proposed method is stored within the organization in a structured form that can serve diverse purposes. To reveal the novelty and difference of the proposed method as an alternative to all traditional methods, some recent studies on WUM preprocessing and web analytics are given in \cref{tab:1}.
	
	Despite some existing studies in the literature conducted on the data obtained using open-source web analytics tools, the number of academic publications that will potentially shed light on the data collection methods of these systems is almost non-existent. This study also clearly identifies, in general, how analytical tools perform the phases such as data obtaining, cleaning, and storage, which is seen as unknown territory. Furthermore, since the proposed method works on the server-side, not client-side, there is no need to send separate requests to the server for data collection. Therefore, the proposed system works faster and makes the server less busy.
	
	The proposed method provides a solution that increases the control of organizations over user data and relieves their apprehensions about delegating their authority to another company. Security concerns and data privacy sensitivity are increasingly growing in today's information societies. Research conducted by Aartsen et al. shows that authors who have published articles on web usage mining consider applying various methods and techniques to web security by using WUM to discover anomalous traffic as one of the most crucial future research topics in this field, along with data privacy \cite{ref42}.
	
	The method also guarantees organizations complete control and ownership of their data. The method's ability to allow flexible data use and focus on data security can provide an effective solution for organizations that spend resources building their internal tools. The method may be preferable in this context, especially in finance, telecommunications, health, and education that work with sensitive data.
	
	It becomes easy and possible to obtain usage data from page requests, especially on large web applications where cross subdomain transactions are carried out with this method. Client and application-specific data are stored in relational databases in a form that can be easily used for data mining or business intelligence applications. Nevertheless, depending on the user density of the site, collected data may increase excessively during the long usage period. In this case, storing historical data in another relational database or a multidimensional structure such as OLAP cubes \cite{ref22} should be considered. Organizations can also use this data to create web analytics systems, monitor sessions in real-time, and detect web application errors or inappropriate usage attempts. 
	
	On the other hand, JavaScript-based methods require a second request to the server. In contrast, the method has fewer steps in the data collection due to its server-side operation principle. Data collection is completed sooner than with any other client-side methods since client data is obtained during the request to the page. Also, there is no need to have JavaScript support on the client side. This situation increases the detection accuracy of search engine bots and other web crawlers.
	
	The proposed innovative method successfully performs user and session identification by increasing data accuracy. Thus, it accelerates the WUM process by eliminating the time-consuming pre-processing step and improves the performance of the mining activity by producing more accurate results. Pattern analysis techniques of WUM, such as association, classification, and clustering, can be applied effortlessly to the data gathered by this method. 
	
	Furthermore, these robust data can be used for web analytics extraction, or they can also be presented as input to artificial intelligence methods. More meaningful information can be obtained from this data by applying knowledge discovery or diverse artificial intelligence methods such as machine learning and deep learning.
	
	To sum up, the major contributions of this study are five-fold:
	
	\begin{enumerate}
		\item  An innovative method is proposed that explains how to obtain, fuse, and clean semi-structured data and store the resulting relational data so that web usage data can be used in both web mining and analytics.
		\item  The method is an application-based and server-side alternative to web server logs used as the data source in web usage mining.
		\item  The method can be used to build web analytics infrastructure that guarantees organizations complete control and ownership of their data. 
		\item  From the web analytics perspective, the proposed method eliminates the need to make a second request to the server for user tracking in client-side methods, thus making the server less busy.
		\item  The structured and relational data obtained through the method are suitable for multi-purpose use, such as machine learning, artificial intelligence, and deep learning, apart from their primary purposes.
	\end{enumerate}

	The remainder of this article is organized as follows. \cref{sec:2} presents brief information about web logs, web analytics, and web mining topics to provide a basis for the study and describes difficulties and related works on preprocessing phase of web usage mining. \cref{sec:3} introduces the proposed method in detail, including three-tier organization, data collection, data model, and data cleaning, processing, and storage stages. \cref{sec:4} illustrates with various analyzes the 24-hour data collected as a result of applying the method to a real system, followed by some concluding remarks in \cref{sec:5}.

	\section{Related work}\label{sec:2}
	
	Web usage mining is a process of analyzing web access logs to comprehend the behavior of users and uncover beneficial patterns. The most common data source seen in studies that summarize much research related to WUM preprocessing phase is web server access logs. A significant part of these works has focused on Apache web server logs. Web server logs are the most frequent and intensive environments of analytical effort.

	\subsection{Web user access logs}\label{sec:2.1}
	
	Due to the nature of the web, users move from one page to another by clicking links on websites. Each new page transition is a request to the web server. Log files have been used to keep track of web requests, also called hits, since the WWW appeared, and the first browser Mosaic was released in 1993 \cite{ref13}. There are various methods for collecting and storing user access data. These methods can be grouped into server-level and application-level. 
	
	Server-level logs are access records obtained and stored in the background during the system's operation by servers providing various services such as web, application, or proxy servers. The access log of a web server contains semi-structured data of all HTTP requests made by browsers. One of the most used web servers globally, Apache uses the NCSA's Common Log Format (CLF) by default, though the server administrator can define a specific format.
	
	The default CLF configuration for access logs of Apache web server is as \cite{ref14}:\newline
	
	\textit{LogFormat ``\%h \%l \%u \%t ``\%r'' \%s \%b'' }\\
	
	The meanings of the arguments included here are as follows:\\
	
	\textit{\%h - The client's IP address. }\par
	\textit{\%l - ID of the person requesting the document as determined by HTTP authentication. }\par
	\textit{\%u - The client's ID if the request is authenticated. }\par
	\textit{\%t - The time the request was received. }\par
	\textit{\%r - The request line containing the HTTP method, protocol, and resource path. }\par
	\textit{\%s - Status code that the server sends back to the client (200: successful, 404: not found, etc.). }\par
	\textit{\%b - The size of the requested object. }\\

	Apache also supports the Extended Common Log Format (ECLF) also known as the NCSA Combined Log Format, which is similar to W3C Extended Log Format (ELF). This log format contains two more information, referrer and user agent, additional to CLF \cite{ref14}.
	
	An example of a simple one-line entry in the access log of an Apache web server, a typical ECLF configuration, is as follows:\\
	
	\indent\textit{193.140.253.80 - - [15/Aug/2021:17:30:51 +0300] ``GET /index.php HTTP/1.1'' 200 1246 ``http://www.server.com/'' ``Mozilla/5.0 (X11; Ubuntu; Linux x86_64; rv:15.0) Gecko/ 20100101 Firefox/15.0.1''}\\
	
	Application server logs, which required a particular set of analyses and were obtained from commercial application servers such as Weblogic, WebSphere, or Tomcat, are other examples of server-level logs \cite{ref15}. Proxy servers and other servers that provide various services also commonly keep logs during their operations. On the other hand, obtaining and storing data in the logs collected at the application level are treated as separate stages.
	
	While the data storage phase occurs on the servers (on-site or off-site), the data-obtaining step can be performed via client-side or server-side web technologies. Application-level user access data is commonly obtained using JavaScript (JS) snippets by the client-side page tagging method, and the logs are stored on remote servers off-site \cite{ref13}. However, this method is unsuitable for organizations that prioritize freely applying web mining techniques to their data.
	
	\subsection{Web analytics}\label{sec:2.2}
	
	Web analytics is defined as the "measurement, collection, analysis, and reporting of internet data for understanding and optimizing web usage" by the Web Analytics Association. Web analytics involves accumulating and evaluating large amounts of data to successfully detect and improve on-site and off-site web usage \cite{ref53}.
	
	The web analytics process entails four steps \cite{ref54}:
	
	\begin{itemize}
		\item \textit{Data collection} (number of visitors, duration of stay on the page, etc.)
		\item \textit{Analytics extraction} (processing the data into information by creating metrics)
		\item \textit{Developing KPIs} (using the information from web analytics metrics for business intelligence)
		\item \textit{Formulating online strategy} (e.g., creating online marketing campaigns)
	\end{itemize}

	\subsection{Web mining}\label{sec:2.3}
	
	Data mining (DM), a step of knowledge discovery from databases (KDD), is a process of obtaining useful information in data using various techniques. DM can be applied to diverse data types such as relational, spatial (geo), multimedia or temporal databases, data warehouses, and web data \cite{ref16}. Although there are various application areas of data mining, the type applied to the web field and uses web data as input is called web mining (WM).Web mining is handled in three categories: Web Content Mining, Web Structure Mining, and Web Usage Mining \cite{ref17}.
	
	\begin{itemize}
		\item \textit{Web Content Mining (WCM)} is about mining information from content on web pages such as text, images, and videos \cite{ref18}.
		
		\item \textit{Web Structure Mining (WSM)} refers to producing the structural summary of websites and is interested in the structure of hyperlinks within the pages \cite{ref19}.
		
		\item \textit{Web Usage Mining (WUM)} focuses on discovering and analyzing clickstream patterns or associated data \cite{ref20}.
	\end{itemize}

	\subsection{Web usage mining}\label{sec:2.4}
	
	Web usage mining is the process of searching for valuable patterns in web access logs containing the users' browsing history \cite{ref21}. Several data mining techniques must be applied to clean raw logs and convert them to transaction itemsets or a data cube. The following three stages of the web usage mining process should be performed to extract information from weblogs.
	
	\textit{Preprocessing} is the first stage, where incomplete or inconsistent existing data will be processed according to the needs of the next phase  \cite{ref38}. \textit{Pattern discovery} is the second stage, where various methods and algorithms such as statistics, data mining, machine learning, and pattern recognition are applied to obtain patterns \cite{ref39}. \textit{Pattern Analysis} is the last stage, where the patterns found are analyzed at this final stage and displayed with understandable interpretations and visualizations \cite{ref40}.

	\subsection{Preprocessing phase of WUM}\label{sec:2.5}
	
	Data preprocessing is the first and most complex task in web usage mining. This crucial stage which generally takes more than 60\% of the whole effort, ensures a structural, reliable, and integrated data source for pattern discovery \cite{ref23}\cite{ref24}. 
	
	The data preprocessing phase consists of separate sub-stages:
	
	\begin{itemize}
		\item \textit{Data filtering} is the process of cleaning data to remove entries that are not directed to users, such as static files and search engine robots that crawl the site \cite{ref26}.
		\item \textit{User identification} is the process of identifying individual users based on their IP addresses and sometimes user agents \cite{ref25}.
		\item \textit{Pageview identification} is the process of determining which page file accesses contribute to a single browser display \cite{ref26}.
		\item \textit{Session identification (or sessionization)} is the process of segmenting the user activity record of each user into sessions to represent a single visit to the site \cite{ref27}.
		\item \textit{Path completion} is the process of identifying accesses that are not logged due to using the local cache method or ``back" button that masks user requests \cite{ref28}.
	\end{itemize}

	\subsection{Challenges in WUM preprocessing}\label{sec:2.6}
	
	Web mining activity is faced primarily with technical issues such as inappropriate data. For successful and reliable mining, the data collected should be in the appropriate format. However, these data are often incomplete or unusable, so their accuracy cannot be ensured. The vast majority of mining endeavors are usually spent on improving data quality \cite{ref37}. 
	
	Raw access log files provided by web servers comprise irrelevant elements affecting the accuracy of pattern discovery and analysis. These inconsequential entries include resources, such as image, CSS, or JavaScript files, unrelated to the page's content, failed requests whose HTTP status code differs from 200, or traces left by search engine indexing robots \cite{ref41}. The following samples of file requests and HTTP status codes can be examined for irrelevant records that need to be filtered:\\
	
	\textit{``GET /site HTTP/1.1'' 301}\par
	\textit{``GET /favicon.ico HTTP/1.1'' 200}\par
	\textit{``GET /app/admin/adm.php HTTP/1.1'' 404}\par
	\textit{``GET /app/images/ccc.png HTTP/l.1'' 200}\par
	\textit{``GET /app/css/login.css HTTP/1.1'' 200}\par
	\textit{``GET /website/ HTTP/1.1'' 302}\\

	Identifying users from web server logs is challenging itself. A user identification problem for the ECLF log format can be formulated as follows.
	
	Consider the $\textit{IP = }\mathrm{\{}$\textit{ip${}_{l}$, ip${}_{2}$, ..., ip${}_{n}$}$\mathrm{\}}$ is a set of all IP addresses of users who accessed the website, $\textit{R = }\mathrm{\{}$\textit{r${}_{l}$, r${}_{2}$, ..., r${}_{n}$}$\mathrm{\}}$ denotes the set of all resources of the website, $\textit{B = }\mathrm{\{}$\textit{b${}_{l}$, b${}_{2}$, ..., b${}_{n}$}$\mathrm{\}}$ is a set of all browsers of web user, and $\textit{K = }\mathrm{\{}$\textit{k${}_{1}$, k${}_{2}$, ..., k${}_{n}$}$\mathrm{\}}$ is a set of links outside the website.
	
	A log entry in ECLF can be defined as $\textit{l${}_{i}$ = }\langle$\textit{ ip${}_{i}$; t; d; r${}_{i}$; v; c; s; }[\textit{ref${}_{i}$}]\textit{;} [\textit{b${}_{i}$}]; [\textit{cookies}] $\rangle$  where \textit{ip${}_{i\ }$$\in$}  \textit{IP}, \textit{r${}_{i\ }$$\in$}  \textit{R}, \textit{ref${}_{i\ }$$\in$}  \textit{R} $\cup$ \textit{K, b${}_{i\ }$$\in$} \textit{B, t }represents \textit{request time as }timestamp, \textit{d} represents \textit{request method }i\textbf{\textit{.}}e. \textit{GET/POST,} \textit{v} represents \textit{HTTP version}, \textit{c} represents \textit{HTTP status code},\textit{ }and\textit{ c} represents\textit{ size of transferred bytes}\textbf{\textit{.}} \textit{ref${}_{i}$, b${}_{i}$, }and\textit{ cookies }are customized attributes.
	
	A web server log consists of \textit{L} = $\mathrm{\{}$\textit{l${}_{1,\ }$l${}_{2}$, {\dots}, l${}_{n}$}$\mathrm{\}}$. Cleaned and extracted log can be defined as \textit{CL =} $\mathrm{\{}$\textit{cl${}_{1}$, cl${}_{2}$, ..., cl${}_{n}$}$\mathrm{\}}$ contains relevant entries and \textit{cl${}_{i}$ = }$\langle$\textit{ ip${}_{i}$; t; r${}_{i}$;} [\textit{ref${}_{i}$}]\textit{; }[\textit{b${}_{i}$}] $\rangle$. Suppose $\mathrm{\textit{U = } \{}$\textit{u${}_{1}$,} \textit{u${}_{2}$, ..., u${}_{n}$}$\mathrm{\}}$ is a set of all users accessed the website.
	
	A visit for a user \textit{u${}_{i}$} to the website can be defined as ${\textit{vs${}_{i}$ = $\langle$ u\textit{${}_{i}$}; e\textit{${}_{i\ }$}$\rangle$}}$, where $\textit{e${}_{i}$ = }\langle$\textit{ }(\textit{t${}_{1}$, r${}_{1}$, }[\textit{ref${}_{1}$}]); (\textit{t${}_{2}$, r${}_{2}$, }[\textit{ref${}_{2}$}]); ...; (\textit{t${}_{n}$, r${}_{n}$, }[\textit{ref${}_{n}$}])\textit{ }$\rangle$\textit{ };\textit{ t${}_{i+1}$ }$\mathrm{\ge}$\textit{ t${}_{i}$}. Accordingly, the user identification is a problem formulated as follows: For a server log \textit{L}, pick up the visits $\mathrm{\textit{V} = \{}$\textit{vs${}_{1,\ }$vs${}_{2}$, {\dots}, vs${}_{n}$}$\mathrm{\}}$ of the web site users from the cleaned log \textit{CL }and write \textit{V }into the user activity file later \cite{ref29}\textbf{\textit{.}}

	Session identification is another major problem in the preprocessing of weblogs \cite{ref9}. A unique session for a server-side programming language allows keeping track of all requests made from the same browser until it is closed. The web servers use cookies to track users and adequately conduct the session process. However, the web servers' principle of keeping access logs is based not on sessions but on logging individual requests line by line. 
	
	A complex set of processes is required to identify users and sessions from the web server logs \cite{ref30}. The main reason why this process is troublesome is that there is no other data in the web server logs to distinguish the user or session, apart from the client's IP address and the time of the request. The limited information in the web server logs is insufficient to understand, for instance, that the browser has been closed and reopened. Commonly, a logging solution tracks visitor sessions by attributing all hits from the same IP address and web browser signature to one person. This situation becomes a problem in combination with the dynamic IP distribution, in which ISPs assign different addresses throughout the session \cite{ref32}.
	
	By checking the time elapsed since the last request, it is attempted to determine whether the user has made a request in a separate session. Requests after a certain period are considered a different session, as the user will not stay on a page for more than a specific time. This time is generally accepted in scientific studies as 10 minutes for consecutive pages and 30 minutes for the session duration \cite{ref31}. Another problem is distinguishing sessions that start before midnight and continue the next day \cite{ref32}. 
	
	There are many studies in the literature related to the need for logging, which is as old as the history of software development. Gholamian et al. \cite{ref36} examined the general log structures used in software and analyzed many log applications comparatively. Srivastava et al. \cite{ref10} revealed that the weblogs containing limited access data presented are a source of inspiration for many studies in the literature.
	
	Paredes et al. \cite{ref5} compiled some studies to extract information from such records. Even though it is reported that largely successful results have been achieved in these studies, the processes, quite laborious, are done offline and do not always give accurate results.  Abdalla et al. \cite{ref2} and Deshpande et al. \cite{ref6} mentioned the difficulties experienced in this regard.
	
	Modeling problems caused mainly by existing algorithms have been studied in previous research, and various suggestions have been made for developing more efficient preprocessing methods. Sukumar et al. \cite{ref7} presented sample pseudo-codes for these studies by conveying the preprocessing techniques in the literature. Kundu et al. \cite{ref8} compiled the weblog analysis applications and compared them in their study.
	
	Fatima et al. \cite{ref9} revealed the studies conducted in session identification in their research. Deshpande et al. \cite{ref6} summarized the work done in user identification in their study. Srivastava et al. \cite{ref10} outlined the techniques applied in data preprocessing of the web server log in their study. Svec et al. \cite{ref11} explained the importance of data preprocessing in web usage mining and the effect of errors on data analysis.
	
	Other than WUM purposes, the biggest problem for organizations that desire to obtain information via cloud-based web analytics tools is privacy concerns, justifiably raised as data is held on third-party servers. Quintel et al. \cite{ref12} raised concerns about this issue in their study. \v{C}egan et al. \cite{ref4} proposed an open web analytics platform called Webalyt, designed to provide usability and reliability under high traffic.
	
	Kumar et al. \cite{ref3} presented a comparative analysis of essential web analytics tools in their research, explaining the challenges to web analytics applications. The on-site tools such as Matomo (formerly Piwik), Open Web Analytics (OWA), and Webalyt use JavaScript-based page tagging methods to obtain data like their cloud-based alternatives. A study explaining the data models, data collection and storage methods, and working principles of Piwik and OWA, the most preferred open-source tools in the field, has not been found in the literature.
	
	All these challenges explain the barriers to successful data collection and efficient mining activity. On the other hand, the content providers are interested mainly in the behavior of the visitors to understand trends and causation besides standard system statistics that are interesting to server administrators \cite{ref33}. Many such requirements cannot be met with simple statistical tools and need more sophisticated techniques.

	\section{The proposed method}\label{sec:3}
	
\begin{figure}[h]
		\centering
		\includegraphics[scale=0.72]{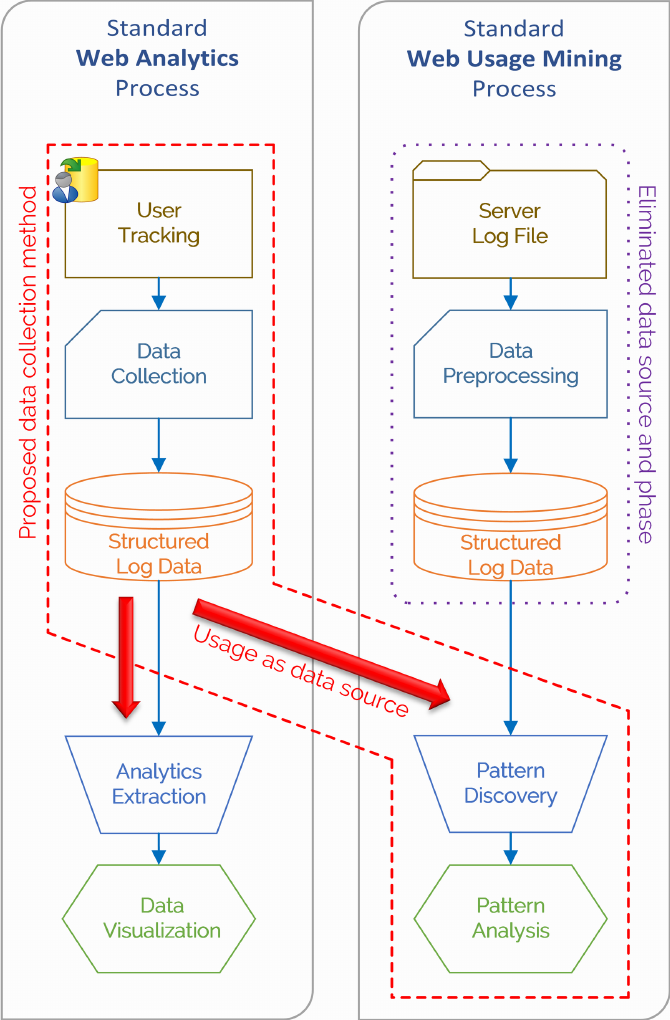}
		\caption{Proposed data collection method and improvement of WUM process}
		\label{fig:1}
	\end{figure}

	Organizations that think the web server logs are insufficient to perform targeted analysis or do not desire to share user access data with others may prefer to keep log data on their servers. The accuracy and integrity of access data are vital to ensure that the web usage mining process yields reliable results. \cref{fig:1} illustrates that the data obtained by a robust collecting method can efficiently be used as a data source for both web analytics extraction and web usage mining.

	It is focused on developing a method for collecting data aimed at WUM more regularly and healthily in this study. The temporal data obtained with the proposed method are stored in a homogeneous structure in the relational database. Using this data also eliminates the need for the preprocessing stage for WUM and presents healthy, clean, and ready-to-use data for the next step directly, pattern discovery.
	
	The proposed data collection method, working on-site, differs from traditional web server logs or client-side forms. The technique relies on using an application-based API to obtain, interpret, format, and store data. The API, developed under the proposed three-tier model, is designed to provide usability and reliability under heavy visitor traffic, adhering to the principles of confidentiality, integrity, and accessibility of information security. The logging API has been implemented into the ``\textit{Campus Automation Web Information System (CAWIS)}" software framework \cite{ref34}, the corporate web application of Sakarya University, and accurate data have been collected. 
	
\begin{figure*}[t]
		\centering
		\includegraphics[scale=0.81]{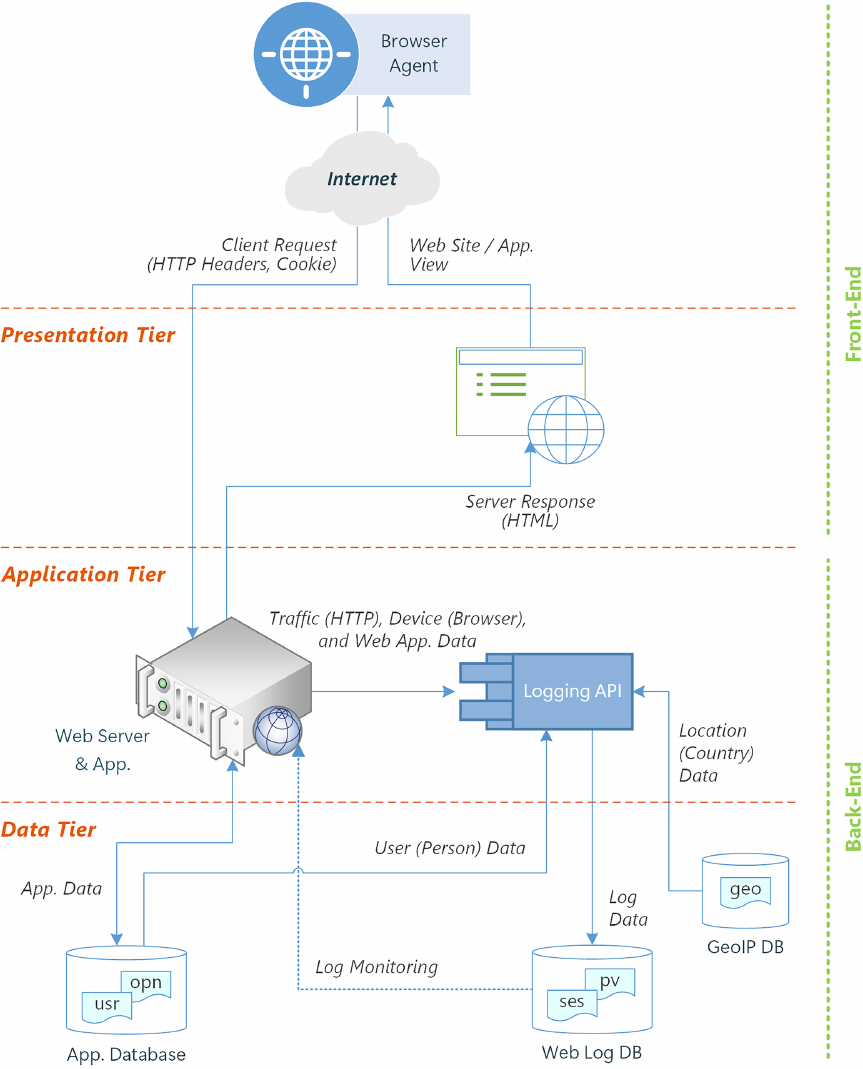}
		\caption{Three-tier organization of the proposed method}
		\label{fig:2}
	\end{figure*}

	The method's three-tier organization and the logging API's positioning are presented in \cref{fig:2}.

	\begin{itemize}
		\item\textit{The presentation tier} includes front-end (client-side) web technologies such as HTML, CSS, and JavaScript. User interfaces of web applications produced on the server side and viewed by the user in the browser are located here.
		
		\item\textit{The application tier} includes the logging API with the application running on the web server. Client and request data from different sources, such as HTTP, network, and web application, are collected and processed here. The application tier is in relationship with both the presentation and data tiers.
		
		\item\textit{The data tier} includes all kinds of database and storage processes. The GeoIP and logging API database, along with application data, is located in this layer. Applications and databases have cross-access according to the model.
	\end{itemize}

	Unlike others, no client-side web technologies have been used for user tracking in this method. The logging API runs server-side and acts as a part of the software framework. Application-level logging is a long-standing approach as a functional characteristic of the web application at a certain level.
	
	The method introduces a more standardized structure instead of the usual or straightforward solutions. Thus, regular data is provided for the pattern discovery stage without data conversion, data cleaning, data filtering, user and session identification, path completion, and data formatting in the preprocessing phase of the web usage mining process.
	
	Moreover, many quantitative and qualitative data not contained by the web server logs are recorded in a structured form for processing later. This data can also be used for various purposes such as instant system monitoring, analytics extraction, and user behavior analysis.

	Another prominent feature of the model is that it can distinguish sessions belonging to dynamic IPs. In the proposed method, sessions are created by the web server, and two different devices are defined as separate sessions, even if they use the same IP address. These sessions are terminated when the user logs out or does not respond to the warning window activated when there is no page activity for a certain period.
	
	In addition, sessions that begin before midnight and continue to the next day are not terminated at the end of the day, unlike other analytics solutions. All these features enable the model to perform much more successful user and session identification than weblogs.

	\subsection{Data collection}\label{sec:3.1}
	
	Enterprise or large-scale web applications are designed generally as a software framework. The log API is built into the framework code to run in the head part. Thus, the API is also called for every page request made.
	
\begin{figure}[h]
		\centering
		\includegraphics*[scale=0.50]{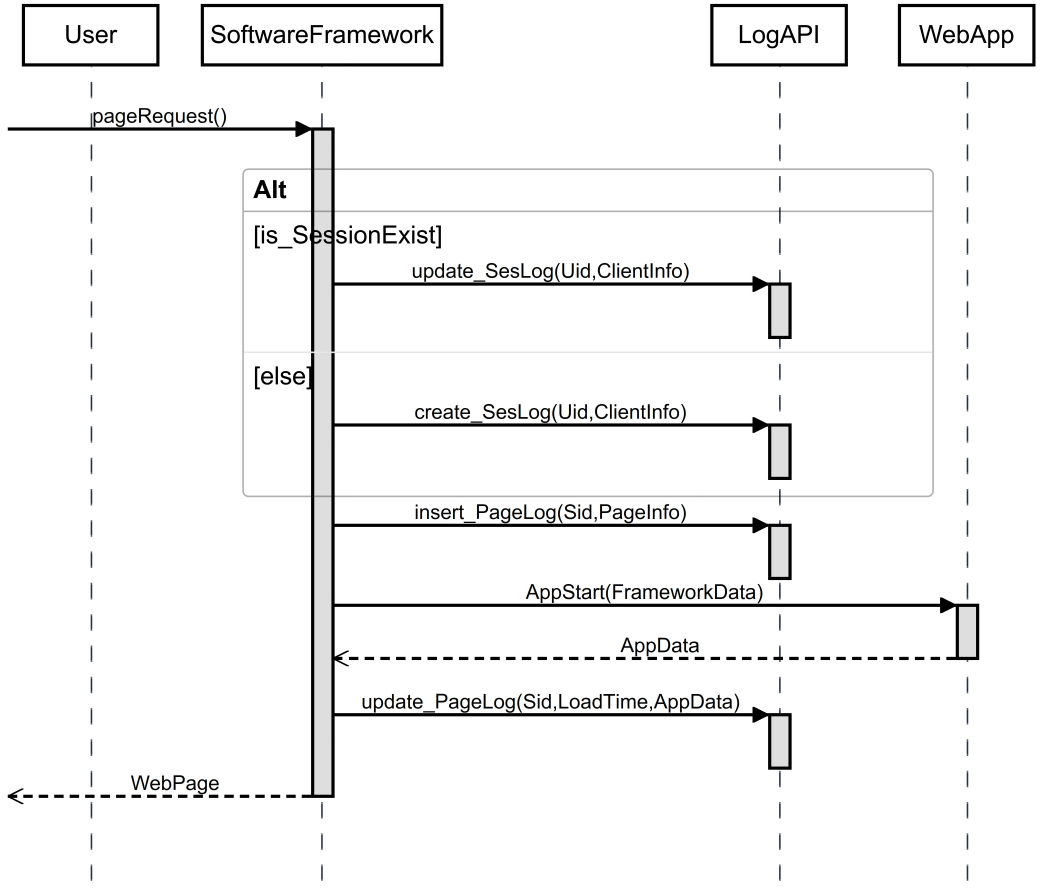}
		\caption{Sequence diagram representation of the way the log API works}
		\label{fig:3}
	\end{figure}

	The working method of the API in the application tier is demonstrated in \cref{fig:3}.
	
	Once the software framework is built appropriately, and the API is integrated correctly, the application developers would have nothing to do with logging. While developers write and publish their application codes, the API always works as part of the layered framework. Before the end of the software framework code, application data about the operations performed on the active page are updated in the page log table. In this way, data for the improvement of the application, such as page load time, delays in database queries, and errors that occur, are obtained.

	The logging API obtains client data from diverse sources such as HTTP, browser-agent, web application, web database, and GeoIP database at the initial stage and combined with data fusion. The data fundamentally belong to one of the four main categorical groups: HTTP request data, network-level data, app-level data, and external data. The logging API collects the major data in \cref{tab:2} with each request.
	
	\begin{table}[h]
		\centering
		\caption{Gathered major data and its sources}
		\label{tab:2}
		\small
		\begin{tabular}{ l  l }
			\toprule
			Data & Source\\
			\midrule
			Client profile/User-agent (browser, OS)  & HTTP request \\
			IP address  & Network protocol  \\
			Referrer (preceding webpage)  & HTTP request  \\
			Referral (channel identification)  & Application  \\
			Geolocation  & External  \\
			Pageview  & HTTP request  \\
			User profile  & Application  \\
			Application-specific data  & Application \\
			Visit or session  & Application  \\
			\hline
		\end{tabular}
	\end{table}
	\smallskip
	
	The data obtained from the HTTP request and network protocol sources contained here are provided to the API by the server-side programming language. The API also accesses application data containing user information shared by the software framework. The client geolocation (country) information is retrieved from the IP data range in the GeoIP table with an SQL query. The logging API obtains detailed data such as browser name, browser version, operating system name, and operating system version using various string processing functions after collecting user agent data containing browser and OS information. The other data in the table are utilized as obtained directly from their source without processing. 
	
	Gathered data are stored in the session and pageview database tables according to the data model after cleaning and processing in the next stage. The database table ${``\textit{open_sessions}"}$ is used to manage sessions and track active users in real-time. A record is created in this table for each user who enters the application. The relevant record is deleted when the user is logged out or the timeout period expires. Thus, it is ensured that the table containing the circulating visitors in the system is constantly up to date.

	The data collection process of the proposed method has \textit{O(N)} linear time complexity depending on the number of requests made. Likewise, the processor power consumed on the server and the data storage size increase proportionally depending on the page requests. The method is in a scalable structure, and its ability to perform operations can be easily increased by providing sufficient resources.

	\subsection{Data model}\label{sec:3.2}
	
\begin{figure*}[t]
		\centering
		\includegraphics[scale=0.675]{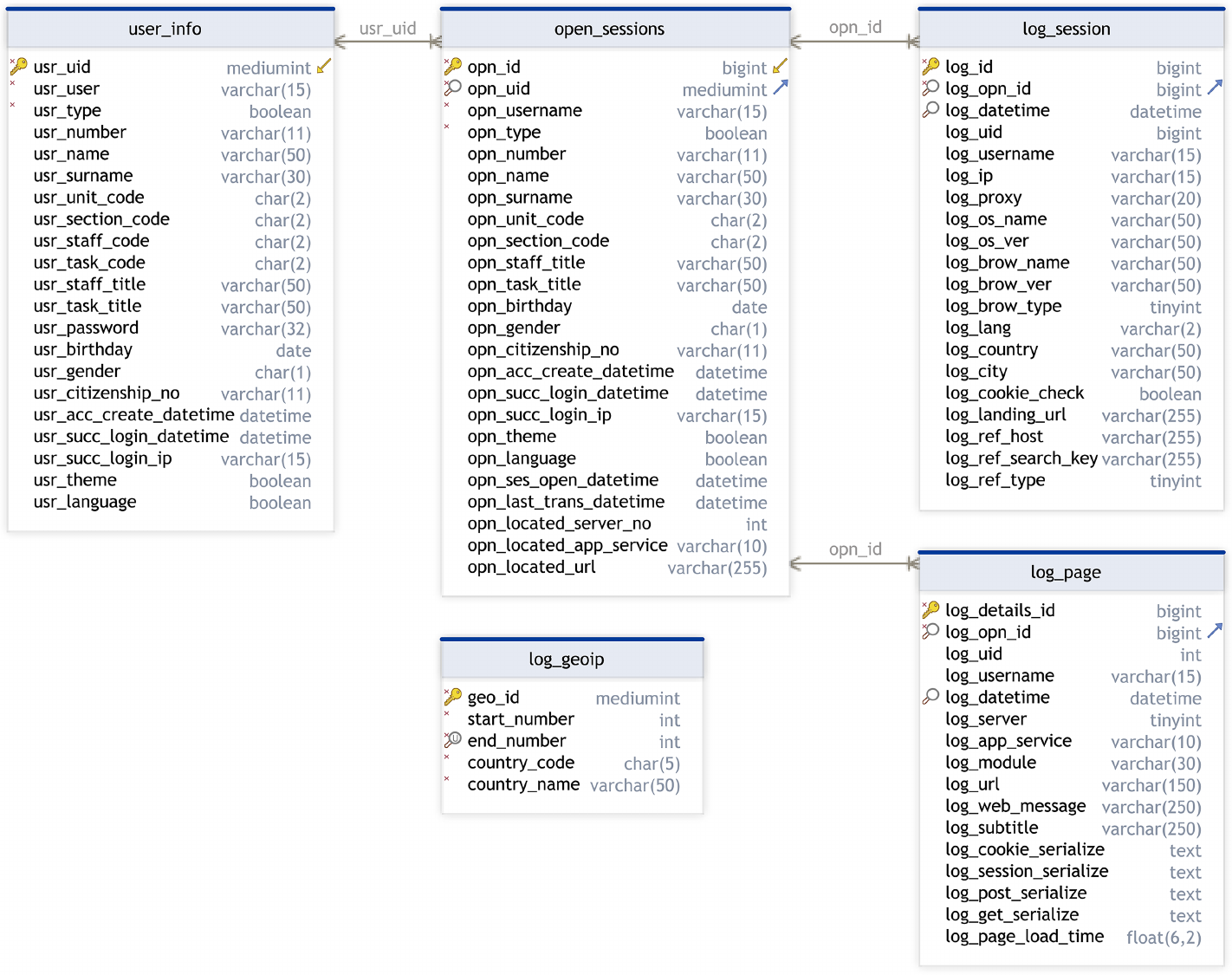}
		\caption{The entity-relationship diagram of the physical data model}
		\label{fig:4}
	\end{figure*}

	Operational data generated during daily transactions in web applications are kept in write-heavy OLTP (Online Transaction Processing) databases, where continuous data input-output operations are performed \cite{ref22}. The records are stored in log tables integrated into the web application database in the proposed method. The entity-relationship diagram (ERD) of the physical data model used in the process is given in \cref{fig:4}.

	\medskip
	The usage purposes of the tables shown in the ER diagram are as follows:
	\medskip
	
	\textit{user\_info:} The user data managed by the web app.
	
	\textit{open\_sessions:} The data of ongoing sessions.
	
	\textit{log\_geoip:} IP ranges data of countries.
	
	\textit{log\_session:} The client data containing session info.
	
	\textit{log\_page:} The request data containing page info.\\
	
	The session management process is carried out by the integrated operation of the application and web server. The first time a request is made to the website, a new alphanumeric session variable is generated automatically for the client by the server-side programming language. On the other hand, the web application uses the table ${``\textit{open_sessions}"}$ to manage ongoing sessions.
	
	All session and page data in log tables are associated through a foreign key named ``\textit{opn\_id}", which consists of an auto-incrementing number for each record in this table. User identification is performed by the API again, which can also access application data.

	\subsection{Data cleaning, processing, and storage}\label{sec:3.3}

	Alongside collecting data, the logging API also carries out the functions of cleaning and storing data. First, all the obtained data are converted into a form that would be stored in the database after cleaning and formatting by data type. Meanwhile, some information such as browser data and request type is encoded. In addition, the country information for determining the client's location is obtained from the GeoIP table via the IP address. Then whole structured data associated with the session are inserted into the session and page log tables which have referential integrity, through queries executed by the API.
	
	Let \textit{S = }$\mathrm{\{}$\textit{s${}_{1}$, s${}_{2}$, ..., s${}_{n}$}$\mathrm{\}}$ denotes the set of sessions created for each visitor. The session data of each \textit{s${}_{i}$} in \textit{S} is stored in a certain row in the session log table of the OLTP database. A session consists of visitor data such as username if logged in, operation system, browser agent, IP, referrer, search engine, and first arrival data that do not change as they browse the pages.
	
	Let \textit{P${}_{s}$ = }$\mathrm{\{}$\textit{p${}_{1}$, p${}_{2}$, ..., p${}_{m}$}$\mathrm{\}}$ denotes the set of page requests in a certain session. The request data of each \textit{p${}_{i}$} in a \textit{P${}_{s}$} is stored in a separate row with a unique session ID for a visitor in the page log table. Primary data such as automatically increasing log details id, session id, user id and name, operation time, server id, application service, page/module, and URL are kept in the page log table. The average record size of one tuple in the tables where the data is stored is about 200 bytes for the session, and 1000 bytes for pageview. A sample tuple from table log_page is given in \cref{tab:6}.

	\begin{table*}[h]
		\centering
		\caption{A sample tuple from table log_page}
		\label{tab:6}
		\footnotesize
		\begin{tabular}{m{0.15\textwidth} >{\raggedright\arraybackslash}m{0.60\textwidth} }
			\toprule
			Table Field	& Sample Data\\
			\midrule
			log\_details\_id & 14036 \\ \hline
			log\_opn\_id & 81291 \\ \hline
			log\_uid & 166553 \\ \hline
			log\_username & user9 \\ \hline
			log\_datetime & 2021-09-02 10:12:18 \\ \hline
			log\_date & 2021-09-02 \\ \hline
			log\_server & 16 \\ \hline
			log\_app\_service & gate \\ \hline
			log\_module & info \\ \hline
			log\_url & http://www.gate.sakarya.edu.tr/?page=info \\ \hline
			log\_web\_message & Welcome to WebGate \\ \hline
			log\_subtitle & Info :: You can review your access and security information here \\ \hline
			log\_cookie\_serialize & a:13:\{s:6:"AUTHID"; s:36:"56b01c19-6aa8-497b-9290-ca8ebb9e6bbc"; s:6:"Apache"; s:27:"10.9.54.19.1630566730862646"; s:9:"PHPSESSID"; s:32:"5282b8b77b0b2cbfe2e854b459f31a7b"; s:10:"cookie\_sid"; s:32:"1765c04ed3a62487d1df2790e078a1ad"; s:11:"cookie\_theme"; s:1:"0"; s:10:"cookie\_lang"; s:1:"0"; s:12:"cookie\_limit"; s:2:"15"; \} \\ \hline
			log\_session\_serialize & a:8:\{s:7:"ses\_uid"; s:6:"166553"; s:6:"ses\_id"; s:5:"81291"; s:8:"ses\_theme"; s:1:"0"; s:8:"ses\_lang"; s:1:"0"; s:9:"ses\_limit"; s:2:"15"; s:12:"ses\_security"; N; s:9:"ses\_child"; b:0; s:8:"ses\_error"; s:0:""; \} \\ \hline
			log\_post\_serialize & a:0:\{\} \\ \hline
			log\_get\_serialize & a:1:\{s:4:"page";s:4:"info";\} \\ \hline
			log\_page\_load\_time & 0,0266 \\ \hline
		\end{tabular}
	\end{table*}
	
	This method has been preferred since at least one data in the table row changes on each page requested. Thus, the user and session identification would be ensured with one hundred percent success. Although the access records in the page log table seem to be in an intermixed order regarding their relationship to sessions, they are stored as a time series in order of requests. However, this situation does not prevent obtaining information about the sessions, and the pages visited in the relevant session via miscellaneous database queries to be made.

	The relation algebra formulation in \cref{eq:1} can be used to obtain the data of pages navigated in separate sessions from log database tables as an SQL inner join.
	
	\begin{equation} \label{eq:1}
		\rho_slog\_session\Join_{s.opn\_id = p.opn\_id}\ \rho_plog\_page
		\medskip
	\end{equation}
	
	In addition, the length of each session can be obtained by summing the visitor's spent time on the pages for that session. Since the requested time \textit{R} is available for each page in the table, the page dwelling time of the \textit{p${}_{th}$} page is calculated by subtracting the requested time of the \textit{p${}_{th}$} page \textit{R${}_{p}$} from the requested time of the \textit{p+1}th page \textit{R${}_{p+1}$} \cite{ref35}.
	
	Accordingly, the dwell time of a session \textit{D${}_{s}$} is calculated using the time scale by the sum of the dwell time of \textit{m} pages as given in \cref{eq:2}.
	
	\begin{equation} \label{eq:2}
		D_s=\sum^m_{p=1}{\left(R_{p+1}-\ R_p\right)}
		\medskip
	\end{equation}
	
	The number of pageviews per session is deliberated according to \cref{eq:3}.
	
	\begin{equation} \label{eq:3}
		P_{ps}=\ \frac{\sum{P}}{m}
		\medskip
	\end{equation}

	The page log table also contains operational data such as application title, application message, cookie, session, post, get, and page load time to ensure information to application developers for software quality improvement and error detection purposes. The web administrator can use these database tables to monitor system usage and track real-time analysis information. In particular, obtaining page production times plays an essential role in web developers being able to detect database-induced slowness.

	\section{Experimental results}\label{sec:4}
	
	No hypothetical or synthetic data has been used in this study. On the contrary, the analyses have been based on the actual data obtained from implementing the method in a real corporate web application. To demonstrate the meaningfulness of collected data by the proposed method, 24-hour data for Friday, March 16, 2018, were examined. Analyzed one-day sample data includes 22,104 sessions and 161,672 pageviews. 
	
	The data collected in the relational database could be analyzed with complex queries generated. This data can also readily adapt to any WUM technique. The data about page requests and sessions allow analysis such as site usages, user types and genders, devices, IPs and proxies, referrer types and hosts, languages and countries, search engines, and keywords. In addition, daily, weekly, monthly, annual, and long-term analytical extractions or comparative analyses can be made over the accumulated data.
	
	\begin{figure*}[t]
		\centering
		\includegraphics[scale=0.95]{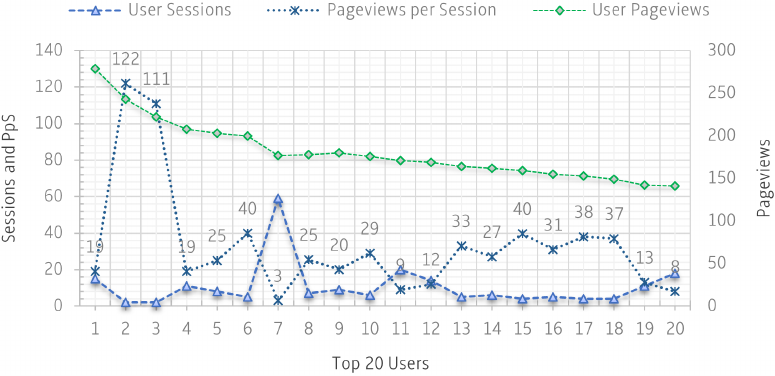}
		\caption{Top 20 users' numbers of pageviews and sessions by the most pageviews in all sessions}
		\label{fig:5}
	\end{figure*}
	
	Depending on the goals and objectives, this data can also be used for anomaly detection and system load estimation. The soundness of the methodology has been assessed through the analysis of the collected actual data. As a result of implementing the method, besides yielding consistent findings, inspiring consequences have also been reached for future activities. Some analyses created to reveal the significance and accuracy of the gathered data are presented in the following subsections. Much more analysis than explained in this section can be easily performed using the data obtained.
	
	\subsection{Analysis of usage}\label{sec:4.1}

	The top 20 users' numbers of pageviews and sessions by the most pageviews in all sessions are shown in \cref{fig:5}. It can be determined how much the users use the system or keep it busy in a sense with this analysis. A similar analysis is the pageviews of the top 20 users by most pageviews in a session. Very high pageviews in a single session can indicate dangerous user action and can be used to detect anomalies.
	
	The number of pages visited by visitors and registered users in the system provides important information about how the system is used. \cref{tab:3} shows that the visitor types and usage count are classified into five groups by frequency of pageviews.

	\begin{table}[h]
		\centering
		\caption{Visitor types and usage count classified into five groups}
		\label{tab:3}
		\small
		\begin{tabular}{ l l r }
			\toprule
			{Visitor Type} & {Pageviews} & {Frequency} \\
			\midrule\noalign{\smallskip}
			\multirow{5}{*}{Guests}
			& 1-3 pages & 20,721 \\
			& 4-10 pages & 1,195 \\
			& 11-30 pages & 62 \\
			& 31-100 pages & 3 \\
			& 101+ pages & 1 \\
			\midrule\noalign{\smallskip}
			\multirow{5}{*}{Users} 
			& 1-3 pages & 5,619 \\
			& 4-10 pages & 9,298 \\
			& 11-30 pages & 2,262 \\
			& 31-100 pages & 282 \\
			& 101+ pages & 13 \\
			\bottomrule
		\end{tabular}
	\end{table}

	This analysis is essential for determining the bounce rate of registered users and visitors and revealing their page browsing habits. The behavior of users browsing more than 100 pages in one session is worth considering separately. In general, the concept of ``visitor" (sometimes ``user") refers to everyone in the system, while where user type is distinguished, ``user" refers to logged-in users, and ``guest" refers to non-logged-in users.

	\begin{table*}[t]
		\centering
		\caption{Number of users, sessions, pageviews, and total session durations by user type and gender}
		\label{tab:4}
		\footnotesize
		\newcolumntype{R}[1]{>{\raggedleft\let\newline\\\arraybackslash}m{#1}}%
		\begin{tabular}{ l l R{1.2cm} r r R{1.6cm} R{1.4cm} R{1.4cm} R{1.4cm} }
			\toprule
			{User Type} & {Gender} & {Number of Users} & {Sessions} & {Pageviews} & {Pageviews per Session} & {Tot. Ses. Dur.(s)} & {Tot. Ses. Dur.(m)} & {Tot. Ses. Dur.(h)}\\
			\midrule\noalign{\smallskip}
			Guest & N/A & 5,343 & 4,655 & 9,006 & 1.93 & - & - & - \\
			Academic Staff & Male & 678 & 2,355 & 28,893 & 12.27 & 778,495 & 12,975 & 216.2 \\
			& Female & 261 & 966 & 12,413 & 12.85 & 363,953 & 6,066 & 101.1 \\
			Administrative Staff & Male & 217 & 548 & 5,606 & 10.23 & 123,551 & 2,059 & 34.3 \\
			& Female & 74 & 205 & 2,193 & 10.70 & 79,217 & 1,320 & 22.0 \\
			Contracted Staff & Male & 14 & 44 & 479 & 10.89 & 24,864 & 414 & 6.9 \\
			& Female & 7 & 16 & 99 & 6.19 & 2,918 & 49 & 0.8 \\
			Retired Staff & Male & 24 & 47 & 438 & 9.32 & 7,874 & 131 & 2.2 \\
			& Female & 4 & 6 & 67 & 11.17 & 5,103 & 85 & 1.4 \\
			Lecturer (Non-Signed) & Male & 11 & 24 & 309 & 12.88 & 11,576 & 193 & 3.2 \\
			& Female & 1 & 4 & 44 & 11.00 & 2,579 & 43 & 0.7 \\
			Student & Male & 5,259 & 7,284 & 57,596 & 7.91 & 1,383,957 & 23,066 & 384.4 \\
			& Female & 4,165 & 5,747 & 42,479 & 7.39 & 950,976 & 15,850 & 264.2 \\
			Graduate & Male & 50 & 81 & 706 & 8.72 & 16,180 & 270 & 4.5 \\
			& Female & 32 & 46 & 445 & 9.67 & 18,054 & 301 & 5.0 \\
			Unit/Mission & N/A & 41 & 76 & 899 & 11.83 & 35,658 & 594 & 9.9 \\
			\noalign{\smallskip}\doubleRulee
			{Total / Average} & & {16,181} & {22,104} & {161,672} & {7.31} & {3,804,955} & {63,416} & {1,057} \\
			\bottomrule
		\end{tabular}
	\end{table*}

	\subsection{Analysis of user types and gender data}\label{sec:4.2}
	
	The number of users, sessions, pageviews, pageviews per session, and total session durations as seconds, minutes, and hours by user type and gender are given in \cref{tab:4}. In addition, average session and page durations can be readily revealed, and evaluations can be made on user behaviors.

	\subsection{Analysis of hourly request data}\label{sec:4.3}
	
\begin{figure}[H]
		\centering
		\includegraphics[scale=0.92]{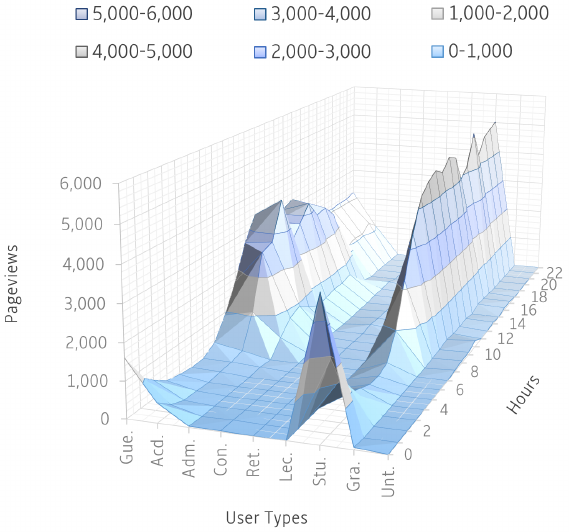}
		\caption{Hourly user numbers of pageviews by user types}
		\label{fig:6}
	\end{figure}

	It is possible to obtain the hourly usage of the system as an OLAP data cube according to user types and genders. These multi-dimensional data are demonstrated in \cref{fig:6} as an instance. Besides providing information about the server and system density, this information is also meaningful for attack and anomaly detection.
	
	\subsection{Analysis of device and browser agent data}\label{sec:4.4}
	
\begin{figure}[h]
		\begin{adjustwidth}{-0.5cm}{0cm}
			\centering
			\par\bigskip
			\begin{subfigure}[t]{0.325\linewidth}
				\includegraphics[scale=0.83]{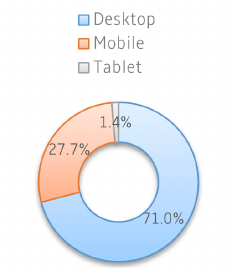}
				\subcaption{}\label{fig:7a}
			\end{subfigure}
			\begin{subfigure}[t]{.325\linewidth}
				\includegraphics[scale=0.83]{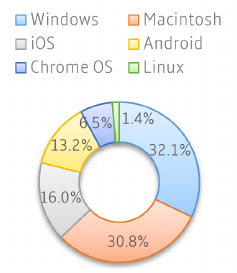}
				\subcaption{}\label{fig:7b}
			\end{subfigure}
			\begin{subfigure}[t]{.325\linewidth}
				\includegraphics[scale=0.83]{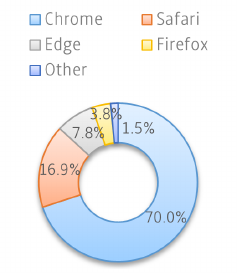}
				\subcaption{}\label{fig:7c}
			\end{subfigure}
			\par\bigskip
			\begin{subfigure}[t]{.325\linewidth}
				\includegraphics[scale=0.83]{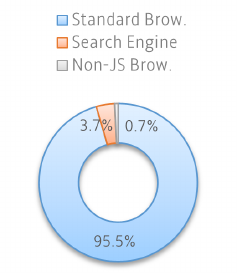}
				\subcaption{}\label{fig:7d}
			\end{subfigure}
			\begin{subfigure}[t]{.325\linewidth}
				\includegraphics[scale=0.83]{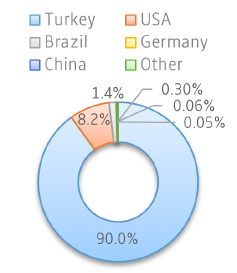}
				\subcaption{}\label{fig:7e}
			\end{subfigure}
			\begin{subfigure}[t]{.325\linewidth}
				\includegraphics[scale=0.83]{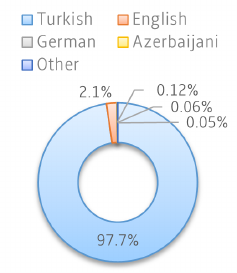}
				\subcaption{}\label{fig:7f}
			\end{subfigure}
			\par\bigskip
		\end{adjustwidth}
		\caption{Technical analysis of client-oriented device and browser agent data}
		\label{fig:7}
	\end{figure}

	Technical analysis of client-oriented device and browser agent data is presented in \cref{fig:7}. Such technical analyses are critical regarding site accessibility, supportability, and usability requirements. The distributions related to used device types in \cref{sub@fig:7a}, the operating systems in \cref{sub@fig:7b}, the browser types in \cref{sub@fig:7c}, and the operating systems in \cref{sub@fig:7d} were obtained by analyzing clients accessing the system. 
	
	In the same way, the countries where users come from in \cref{sub@fig:7e} and the browser languages in \cref{sub@fig:7f} are presented as ratios. The information, such as the browser distribution or language, reveals the necessity of design and script tests accomplished on different browsers after the website is coded. Considering that the websites are the doors of organizations opening to the world, this information becomes even more meaningful.

	\subsection{Analysis of IP data}\label{sec:4.5}
	
	The number of visits and pageviews of the top 15 IP addresses (hidden) sorted by the number of sessions opened is given in \cref{tab:5}. The same analysis can also be obtained according to the proxy IPs. These data are essential for the detection of intensive or dangerous uses.
	
	\begin{table}[H]
		\centering
		\caption{Top 15 IP addresses' (hidden) numbers of visits and pageviews}
		\label{tab:5}
		\small
		\newcolumntype{R}[1]{>{\raggedleft\let\newline\\\arraybackslash}m{#1}}%
		\begin{tabular}{ l r r R{1.6cm} }
			\toprule
			{IP Addresses} & {Sessions} & {Pageviews} & {Pageviews per Session} \\
			\midrule\noalign{\smallskip}
			IP1 & 416 & 933 & 2.24 \\
			IP2 & 134 & 706 & 5.27 \\
			IP3 & 108 & 163 & 1.51 \\
			IP4 & 104 & 642 & 6.17 \\
			IP5 & 102 & 556 & 5.45 \\
			IP6 & 98 & 474 & 4.84 \\
			IP7 & 85 & 416 & 4.89 \\
			IP8 & 82 & 501 & 6.11 \\
			IP9 & 81 & 455 & 5.62 \\
			IP10 & 78 & 422 & 5.41 \\
			IP11 & 76 & 303 & 3.99 \\
			IP12 & 69 & 464 & 6.72 \\
			IP13 & 66 & 368 & 5.58 \\
			IP14 & 65 & 408 & 6.28 \\
			IP15 & 63 & 391 & 6.21 \\
			\bottomrule
		\end{tabular}
	\end{table}

	\subsection{Analysis of search engine data}\label{sec:4.6}
	
	The frequency of search engines used while accessing the website by visitors is shown in \cref{sub@fig:8a}. Also, the distribution of search keywords used by visitors to access the site via search engines is given in \cref{sub@fig:8b}.
	
	\begin{figure}[H]
		\centering
		\begin{subfigure}{\columnwidth}
			\centering
			\includegraphics[scale=0.91]{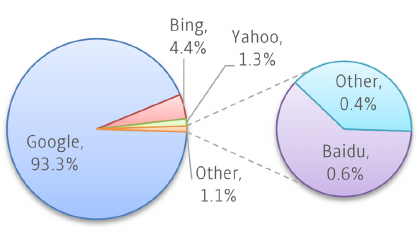}
			\subcaption{}\label{fig:8a}
		\end{subfigure}
		\par\bigskip
		\begin{subfigure}{\columnwidth}
			\centering
			\includegraphics[scale=0.94]{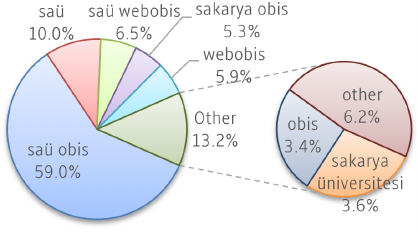}
			\subcaption{}\label{fig:8b}
		\end{subfigure}
		\caption{Frequency of search engines and search keywords}
		\label{fig:8}
	\end{figure}
	
	Commercial sites should consider this data due to their efforts to detect sponsored links accessed from search engines or understand the method of users finding the site. Evaluating which expressions would be more efficient to pay a sponsor link fee or reviewing the site organization in the light of this data are crucial for effective site structuring.

	\section{Conclusion and future work}\label{sec:5}
	
	A data collection method that works at the application level has been presented in this study to generate a homogeneous data infrastructure suitable for analyzing the behaviors exhibited by visitors while browsing the websites. The proposed method also eliminates the obligation to share data with third-party companies today, where data privacy and sovereignty concerns are increasing. The study also focuses on how the relational data collected using the proposed model will improve the whole process by eliminating the need for the preprocessing stage of web usage mining. 
	
	The proposed method was implemented in the corporate web application of a university, which had a user potential that was well above the average and an extensive visitor base. The logging API, developed according to the method, successfully fulfilled the desired aims. The results show that the method is highly efficient in data collection and allows this data to be evaluated in various ways. In this context, examining the collected 24-hour data will be sufficient to show that the method works well and is usable. Further, the proposed method has reached a substantial number of log records, such as 22 million sessions and 142 million page views in a year. The true success of this method is based on the robustness of the methodology applied to collect data. Completely clean and hassle-free, the data could be presented as input to artificial intelligence methods such as machine learning or deep learning for knowledge discovery.
	
	The three-tier model proposed in the method has a scalable structure with its functions, the database used, the smoothness of the code base, and the possibility of testing and maintenance. The model is designed to respond to increased requests as long as there is no bottleneck in the server's processing power, bandwidth, and disk space. There is no theoretical limit to the data size that the method can handle, except for the database disk space. In case of an increase in requests or data for web applications of different sizes and intensities, additional resources can be added when necessary, and the increased workload can be met without any disruption. 
	
	Although the model has been applied to an RDBMS, the fact that any database management system or NoSQL is being used on the data tier does not change the originality of the proposed method. Requirements such as speed and capacity are issues that need to be thought of independently of the model and focused on for institutions and organizations that will use this method.
	
	Even though a university corporate web application is given as an example in the application layer, the model could be adapted to many different platforms, such as commercial or banking sites, large or small. In case of increasing requests or data for web applications of different sizes and densities, additional resources can be added if necessary, and increased workload can be met without interruption.
	
	It is as essential to present data appropriately when necessary as collecting, processing, and storing them. Extracted analytical information should be accessible and available to authorized persons when needed. Although not included in this paper, the model outputs can be presented in a user-friendly format with various visualizations and dashboards at the presentation tier.
	
	A known drawback of this method is the inability to collect data such as screen resolution that can only be obtained with client-side web technologies. This feature is naturally available in analytics tools that gather data via JavaScript. Even though it is not a requisite, such client data obtained using an intermediary JavaScript snippet can be transmitted to the API as application-specific data. Nonetheless, AJAX technology allows sending and receiving data in the background between the web application and the server, and it has native compatibility with the API. In addition, applying the method requires programming skills, database knowledge, time, and effort.
	
	Merely the stage of collecting web usage data is focused on, and the benefits of collecting data via the proposed method are discussed in this study. However, the accumulated information over time causes the used physical size to increase constantly and the database tables to swell excessively. Moreover, the operational database, writing-heavy, is not suitable for performing queries on it and executing mining operations.
	
	The accumulated data from the previous day should be turned into statistics information by analytics extraction every midnight and should be transferred to the data warehouse by ETL process for future use. Thus, the analytical information obtained can be examined hourly, daily, weekly, monthly, or yearly, and correlations between them could be analyzed. In addition, mining operations could be performed smoothly on long-term data and reading-heavy databases. Future work will focus on extracting analytical information from daily log data and storing batch data.
	
	\section*{Declaration of competing interest}\label{declare}
	The authors declare that they have no known competing financial interests or personal relationships that could have appeared to influence the work reported in this paper.
	
	\section*{Acknowledgements}\label{ack}
	This study has been supported within the doctoral thesis support project (2010-50-02-024) by the Sakarya University Scientific Research Projects Commission. The corporate web application ``CAWIS'' in the structure of the software framework used to collect data in the study has also been supported by the same organization within the research support project (2007-01-10-001).

	\bibliographystyle{elsarticle-num}

\begin{thebibliography}{10}
\expandafter\ifx\csname url\endcsname\relax
  \def\url#1{\texttt{#1}}\fi
\expandafter\ifx\csname urlprefix\endcsname\relax\def\urlprefix{URL: }\fi
\expandafter\ifx\csname href\endcsname\relax
  \def\href#1#2{#2} \def\path#1{#1}\fi

\bibitem{ref1}
V.~Jain, K.~Kashyap, An efficient algorithm for web log data preprocessing, in:
  Machine Vision and Augmented Intelligence---Theory and Applications, Springer
  Singapore, Singapore, 2021, p. 505–514.
\newblock doi: 10.1007/978-981-16-5078-9_41.

\bibitem{ref2}
A.~Abdalla, T.~Ahmed, M.~Seliaman, Web usage mining and the challenge of big
  data: A review of emerging tools and techniques, in: I.R.M. Association
  (Ed.), Big Data: Concepts, Methodologies, Tools, and Applications, vol.~6,
  IGI Global, 2016, Ch.~42, p. 899–928.
\newblock doi: 10.4018/978-1-4666-9840-6.ch042.

\bibitem{ref3}
V.~Kumar, G.~Ogunmola, Web analytics for knowledge creation: A systematic
  review of tools, techniques, and practices, International Journal of Cyber
  Behavior, Psychology and Learning (IJCBPL) 10~(1) (2020) 1–14.
\newblock doi: 10.4018/IJCBPL.2020010101.

\bibitem{ref4}
L.~{\v{C}}egan, P.~Filip, Webalyt: Open web analytics platform, in: 2017 27th
  International Conference Radioelektronika, IEEE, 2017, p. 1–5.
\newblock doi: 10.1109/RADIOELEK.2017.7937605.

\bibitem{ref43}
Y.~Tao, S.~Guo, C.~Shi, D.~Chu, User behavior analysis by cross-domain log data
  fusion, IEEE Access 8 (2019) 400--406.
\newblock doi: 10.1109/ACCESS.2019.2961769.

\bibitem{ref44}
S.A. Ehikioya, S.~Lu, A path analysis model for effective e-commerce
  transactions, African Journal of Computing and ICT 12~(2) (2019) 55--71.

\bibitem{ref45}
R.~Roy, G.A. Rao, Survey on pre-processing web log files in web usage mining,
  International Journal of Advanced Science and Technology 29~(3 Special Issue)
  (2020) 682--691.

\bibitem{ref46}
K.K. Ibrahim, A.J. Obaid, Web mining techniques and technologies: A landscape
  view, Journal of Physics: Conference Series 1879~(3) (2021) 032125.
\newblock doi: 10.1088/1742-6596/1879/3/032125.

\bibitem{ref47}
M.~Srivastava, A.K. Srivastava, R.~Garg, P.~Mishra, Performance evaluation of
  the mapreduce-based parallel data preprocessing algorithm in web usage mining
  with robot detection approaches, IETE Technical Review (2021) 1--15doi:
  10.1080/02564602.2021.1918584.

\bibitem{ref48}
M.A. Bayir, I.H. Toroslu, Maximal paths recipe for constructing web user
  sessions, World Wide Web (2022) 1--31doi: 10.1007/s11280-022-01024-3.

\bibitem{ref49}
M.~Munk, L.~Benko, Using entropy in web usage data preprocessing, Entropy
  20~(1) (2018) 67.
\newblock doi: 10.3390/e20010067.

\bibitem{ref50}
M.~Srivastava, R.~Garg, P.~Mishra, A mapreduce-based user identification
  algorithm in web usage mining, International Journal of Information
  Technology and Web Engineering (IJITWE) 13~(2) (2018) 11--23.
\newblock doi: 10.4018/IJITWE.2018040102.

\bibitem{ref51}
S.~Knight-Davis, Using awstats to analyze logs from ezproxy and from the public
  opac logs, in: Spring Forum: Collection Management and Technical Services
  Committees, 2017, p. 228.

\bibitem{ref52}
J.~Gamalielsson, B.~Lundell, S.~Butler, C.~Brax, T.~Persson, A.~Mattsson,
  T.~Gustavsson, J.~Feist, E.~L{\"o}nroth, Towards open government through open
  source software for web analytics: The case of matomo, JeDEM-eJournal of
  eDemocracy and Open Government 13~(2) (2021) 133--153.
\newblock doi: 10.29379/jedem.v13i2.650.

\bibitem{ref42}
B.~Aartsen, O.F. El-Gayar, C.~Noteboom, A systematic review of web usage mining
  techniques and future research options, in: MWAIS 2020 Proceedings, vol.~25,
  MWAIS, 2020, pp. 1--6.

\bibitem{ref22}
B.~Milosevic, D.~Regodic, V.~Saso, Big data management processes in business
  intelligence systems, in: Economic and Social Development: Book of
  Proceedings, Varazdin Development and Entrepreneurship Agency (VADEA), 2021,
  pp. 182--192.

\bibitem{ref13}
G.~Zheng, S.~Peltsverger, Web analytics overview, in: Encyclopedia of
  Information Science and Technology, Third Edition, IGI Global, 2015, p.
  7674–7683.
\newblock doi: 10.4018/978-1-4666-5888-2.ch756.

\bibitem{ref14}
M.~Srivastava, R.~Garg, P.K. Mishra, Analysis of data extraction and data
  cleaning in web usage mining, in: Proceedings of the 2015 International
  Conference on Advanced Research in Computer Science Engineering \& Technology
  (ICARCSET 2015), Association for Computing Machinery, 2015, p. 1–6.
\newblock doi: 10.1145/2743065.2743078.

\bibitem{ref15}
S.~Mishra, S.~Srivastava, Web development frameworks and its performance
  analysis—a review, Smart Computing (2021) 337–343doi:
  10.1201/9781003167488-39.

\bibitem{ref53}
B.~Clifton, Advanced web metrics with google analytics, in: Advanced Web
  Metrics with Google Analytics, 3rd Edition, John Wiley \& Sons: Indianapolis,
  IN, USA, 2012, pp. 3--5.

\bibitem{ref54}
I.~Onder, A.~Berbekova, Web analytics: more than website performance
  evaluation?, International Journal of Tourism Citiesdoi:
  10.1108/IJTC-03-2021-0039.

\bibitem{ref16}
G.~Reddy, A review of data warehouses multidimensional model and data mining,
  Information Technology in Industry 9~(3) (2021) 310–320.

\bibitem{ref17}
P.~Shah, H.~Pandit, A review: Web content mining techniques, Data Engineering
  for Smart Systems (2022) 159–172doi: 10.1007/978-981-16-2641-8_15.

\bibitem{ref18}
P.~Shah, H.B. Pandit, A review: Web content mining techniques, Data Engineering
  for Smart Systems (2022) 159--172doi: 10.1007/978-981-16-2641-8_15.

\bibitem{ref19}
N.~Tyagi, S.K. Gupta, Web structure mining algorithms: A survey, in: Big Data
  Analytics, Springer, 2018, pp. 305--317.
\newblock doi: 10.1007/978-981-10-6620-7_30.

\bibitem{ref20}
Z.Y. Lim, L.Y. Ong, M.C. Leow, A review on clustering techniques: Creating
  better user experience for online roadshow, Future Internet 13~(9) (2021)
  233.
\newblock doi: 10.3390/fi13090233.

\bibitem{ref21}
R.~Das, I.~Turkoglu, Creating meaningful data from web logs for improving the
  impressiveness of a website by using path analysis method, Expert Systems
  with Applications 36~(3) (2009) 6635–6644.
\newblock doi: 10.1016/j.eswa.2008.08.067.

\bibitem{ref38}
M.~Manchanda, N.~Gupta, Web usage mining: Dynamic methodology to preprocessing
  web logs, HELIX 8~(5) (2018) 3810--3815.
\newblock doi: 10.29042/2018-3810-3815.

\bibitem{ref39}
N.~Jokar, A.R. Honarvar, S.~Aghamirzadeh, K.~Esfandiari, Web mining and web
  usage mining techniques, Bulletin de la Soci{\'e}t{\'e} des Sciences de
  Li{\`e}ge 85~(1) (2016) 321--328.
\newblock doi: 10.25518/0037-9565.5371.

\bibitem{ref40}
S.~Kumar, R.~Kumar, A study on different aspects of web mining and research
  issues, in: IOP Conference Series: Materials Science and Engineering, vol.
  1022, IOP Publishing, 2021, pp. 012--018.
\newblock doi: 10.1088/1757-899X/1022/1/012018.

\bibitem{ref23}
L.~Kewen, Analysis of preprocessing methods for web usage data, in: Proceedings
  of 2012 International Conference on Measurement, Information and Control,
  vol.~1, IEEE, 2012, p. 383–386.
\newblock doi: 10.1109/MIC.2012.6273276.

\bibitem{ref24}
B.~Mobasher, Web mining overview, in: J.~Wang (Ed.), Encyclopedia of Data
  Warehousing and Mining, Second Edition, 2nd Edition, vol.~3, IGI Global,
  2009, Ch. 319, pp. 2085--2089.
\newblock doi: 10.4018/978-1-60566-010-3.ch319.

\bibitem{ref26}
G.~Slanzi, G.~Pizarro, J.D. Vel{\'a}squez, Biometric information fusion for web
  user navigation and preferences analysis: An overview, Information Fusion 38
  (2017) 12--21.
\newblock doi: 10.1016/j.inffus.2017.02.006.

\bibitem{ref25}
R.~Nandal, et~al., A systematic review on data preprocessing and pattern
  discovery of web usage mining, International Journal of Advanced Research in
  Computer Science 9~(2).
\newblock doi: 10.26483/ijarcs.v9i2.5763.

\bibitem{ref27}
B.~Fatima, H.~Ramzan, S.~Asghar, Session identification techniques used in web
  usage mining: a systematic mapping of scholarly literature, Online
  Information Reviewdoi: 10.1108/OIR-08-2015-0274.

\bibitem{ref28}
T.~Joachims, L.~Granka, B.~Pan, H.~Hembrooke, G.~Gay, Accurately interpreting
  clickthrough data as implicit feedback, in: ACM SIGIR Forum, vol.~51, ACM,
  New York, NY, USA, 2017, p. 4–11.
\newblock doi: 10.1145/1076034.1076063.

\bibitem{ref37}
S.~Garc{\'\i}a, J.~Luengo, F.~Herrera, Data preprocessing in data mining,
  vol.~72, Springer, 2015.
\newblock doi: 10.1007/978-3-319-10247-4.

\bibitem{ref41}
N.~Kaur, H.~Aggarwal, A novel semantically-time-referrer based approach of web
  usage mining for improved sessionization in pre-processing of web log,
  International journal of advanced computer science and applications 8~(1).
\newblock doi: 10.14569/IJACSA.2017.080122.

\bibitem{ref29}
M.~Srivastava, R.~Garg, P.~Mishra, A mapreduce-based user identification
  algorithm in web usage mining, International Journal of Information
  Technology and Web Engineering (IJITWE) 13~(2) (2018) 11–23.
\newblock doi: 10.4018/IJITWE.2018040102.

\bibitem{ref9}
B.~Fatima, H.~Ramzan, S.~Asghar, Session identification techniques used in web
  usage mining: A systematic mapping of scholarly literature, Online
  Information Review 40~(7) (2016) 1033–1053.
\newblock doi: 10.1108/OIR-08-2015-0274.

\bibitem{ref30}
M.J.H. Mughal, Data mining: Web data mining techniques, tools and algorithms:
  An overview, International Journal of Advanced Computer Science and
  Applications 9~(6).
\newblock doi: 10.14569/IJACSA.2018.090630.

\bibitem{ref32}
B.~Clifton, Advanced web metrics with Google Analytics, John Wiley \& Sons,
  2012.

\bibitem{ref31}
C.R. Varnagar, N.N. Madhak, T.M. Kodinariya, J.N. Rathod, Web usage mining: A
  review on process, methods and techniques, in: 2013 International Conference
  on Information Communication and Embedded Systems (ICICES), IEEE, 2013, pp.
  40--46.
\newblock doi: 10.1109/ICICES.2013.6508399.

\bibitem{ref36}
S.~Gholamian, P.~Ward, A comprehensive survey of logging in software: From
  logging statements automation to log mining and analysis, arXiv preprint
  arXiv:2110.12489doi: 10.48550/arXiv.2110.12489.

\bibitem{ref10}
M.~Srivastava, A.~Srivastava, R.~Garg, Data preprocessing techniques in web
  usage mining: A literature review, in: Proceedings of International
  Conference on Sustainable Computing in Science, Technology and Management
  (SUSCOM), Amity University Rajasthan, Jaipur-India, 2019, pp. 466--476.
\newblock doi: 10.2139/ssrn.3352357.

\bibitem{ref5}
R.~Paredes, J.~Bolanio, Analyzing logs from proxy server and captive portal
  using k-means clustering algorithm, Middle East Journal of Applied Science \&
  Technology 3~(4) (2020) 10–31.
\newblock doi: 10.4018/IJCBPL.2020010101.

\bibitem{ref6}
D.~Deshpande, S.~Deshpande, V.~Thakare, Web user identification: Analysis of
  heuristic solutions, in: 2018 Second International Conference on Intelligent
  Computing and Control Systems (ICICCS), IEEE, 2018, p. 1790–1795.
\newblock doi: 10.1109/ICCONS.2018.8662893.

\bibitem{ref7}
P.~Sukumar, L.~Robert, S.~Yuvaraj, Review on modern data preprocessing
  techniques in web usage mining (wum), in: 2016 International Conference on
  Computation System and Information Technology for Sustainable Solutions
  (CSITSS), IEEE, 2016, p. 64–69.
\newblock doi: 10.1109/CSITSS.2016.7779441.

\bibitem{ref8}
S.~Kundu, L.~Garg, Web log analyzer tools: A comparative study to analyze user
  behavior, in: 2017 7th International Conference on Cloud Computing, Data
  Science \& Engineering-Confluence, IEEE, 2017, p. 17–24.
\newblock doi: 10.1109/CONFLUENCE.2017.7943117.

\bibitem{ref11}
P.~Svec, L.~Benko, M.~Kadlecik, J.~Kratochvil, M.~Munk, Web usage mining: Data
  pre-processing impact on found knowledge in predictive modelling, Procedia
  Computer Science 171 (2020) 168–178.
\newblock doi: 10.1016/j.procs.2020.04.018.

\bibitem{ref12}
D.~Quintel, R.~Wilson, Analytics and privacy, Information Technology and
  Libraries 39~(3) (2020) 1–11.
\newblock doi: 10.6017/ital.v39i3.12219.

\bibitem{ref33}
O.~Canay, U.~Kocabicak, A new data collection model for information extraction
  from web click logs, in: International Artificial Intelligence and Data
  Processing Symposium (IDAP), Inonu University, Malatya, Turkey, 2016, p.
  489–492.

\bibitem{ref34}
O.~Canay, S.~Meric, H.~Evirgen, M.~Varan, Realization of campus automation web
  information system in context of service unity architecture, in:
  International Symposium on Computing in Science \& Engineering (ISCSE),
  Izmir, Turkey, 2011, p. 173–179.

\bibitem{ref35}
S.~Malarvizhi, B.~Sathiyabhama, Frequent pagesets from web log by enhanced
  weighted association rule mining, Cluster Computing 19~(1) (2016) 269–277.
\newblock doi: 10.1007/s10586-015-0507-z.

\end{thebibliography}

\end{document}